\documentclass[prb,a4paper,twocolumn,floatfix,showpacs,showkeys,amsmath,amssymb,nobibnotes,altaffilletter]{revtex4-1}

\usepackage{graphicx}% Include figure files
\usepackage{pslatex}
\usepackage{xspace}
\usepackage{subfigure}
\usepackage{color}

\newcommand{\PEDOT}{poly(3,4-ethylenedioxythiophene)\xspace}
\newcommand{\PSS}{poly(styrenesulfonate)\xspace}
\newcommand{\degree}{$\,^{\circ}\mathrm{C}$\xspace}
\newcommand{\pht}{poly(3-hexylthiophene-2,5-diyl)\xspace}
\newcommand{\pcbm}{[6,6]-phenyl-C$_{61}$ butyric acid methyl ester\xspace}

\newcommand{\vpos}{$V_\text{POS}$\xspace}

\newcommand{\vcv}{$V_\text{CV}$\xspace}
\newcommand{\vbi}{$V_\text{BI}$\xspace}
\newcommand{\jph}{$J_\text{Ph}$\xspace}
\newcommand{\offset}{$J_\text{Ph}(V_\text{POS})$\xspace}

\newcommand{\jsc} {$J_\text{SC}$\xspace}

\newcommand{\rpp}{$r_\text{PP}$\xspace}

\newcommand{\mutau}{$\mu\tau_f$\xspace}

\begin{document}

\title{Investigation of the Photocurrent in Bulk Heterojunction Solar Cells}

\author{M.~Limpinsel}
\affiliation{Experimental Physics VI, Julius-Maximilians-University of W{\"u}rzburg, D-97074 W{\"u}rzburg}
\author{A.~Wagenpfahl}
\affiliation{Experimental Physics VI, Julius-Maximilians-University of W{\"u}rzburg, D-97074 W{\"u}rzburg}
\author{M.~Mingebach}
\affiliation{Experimental Physics VI, Julius-Maximilians-University of W{\"u}rzburg, D-97074 W{\"u}rzburg}
\author{C.~Deibel}\email{deibel@physik.uni-wuerzburg.de}
\affiliation{Experimental Physics VI, Julius-Maximilians-University of W{\"u}rzburg, D-97074 W{\"u}rzburg}
\author{V.~Dyakonov}
\affiliation{Experimental Physics VI, Julius-Maximilians-University of W{\"u}rzburg, D-97074 W{\"u}rzburg}
\affiliation{Functional Materials for Energy Technology, Bavarian Centre for Applied Energy Research (ZAE Bayern), D-97074 W{\"u}rzburg}

\date{\today}

\begin{abstract}

We investigated the photocurrent in \pht (P3HT):\pcbm (PCBM) solar cells by applying a pulsed measurement technique. For annealed samples, a point of optimal symmetry (POS) with a corresponding voltage \vpos of 0.52--0.64 V could be determined. Based on macroscopic simulations and results from capacitance--voltage measurements, we identify this voltage with flat band conditions in the bulk of the cell, but not the built-in voltage as proposed by [Ooi et al., J. Mater. Chem. 18 (2008) 1644]. We calculated the field dependent polaron pair dissociation after Onsager--Braun and the voltage dependent extraction of charge carriers after Sokel and Hughes with respect to this point of symmetry. Our analysis allows to explain the experimental photocurrent in both forward and reverse directions. Also, we observed a voltage--independent offset of the photocurrent. As this offset is crucial for the device performance, we investigated its dependence on cathode material and thermal treatment. From our considerations we gain new insight into the photocurrent`s voltage dependence and the limitations of device efficiency.

\end{abstract}



\maketitle

\section{Introduction}

Organic solar cells have improved greatly in the last years, reaching 5--6 \% power conversion efficiency (PCE) today.\cite{Park2009} However, the Shockley diode equation cannot explain the voltage dependent photocurrent in organic solar cells based on a physical model. \cite{schilinsky2004} The detailed process leading from photoinduced polaron pairs to extracted charge carriers and external photocurrent needs better understanding to push development of these devices systematically.\cite{deibel2009c} 
The first step in this process is singlet exciton formation upon absorption of a photon, usually on the polymer. Then, the exciton has to diffuse to a polymer-fullerene interface within its lifetime, where, due to the high electron affinity of the fullerene, a polaron pair is created via ultrafast charge-transfer.\cite{sariciftci1992} These polaron pairs are still Coulomb-bound due to a low dielectric constant $\epsilon_r$ of the organic material system (typically 3--4), and have to be dissociated to free polarons. Provided the free polarons arrive at their respective electrodes, the last step is charge extraction.

\begin{figure}[tb]
	\includegraphics[height=7.5cm]{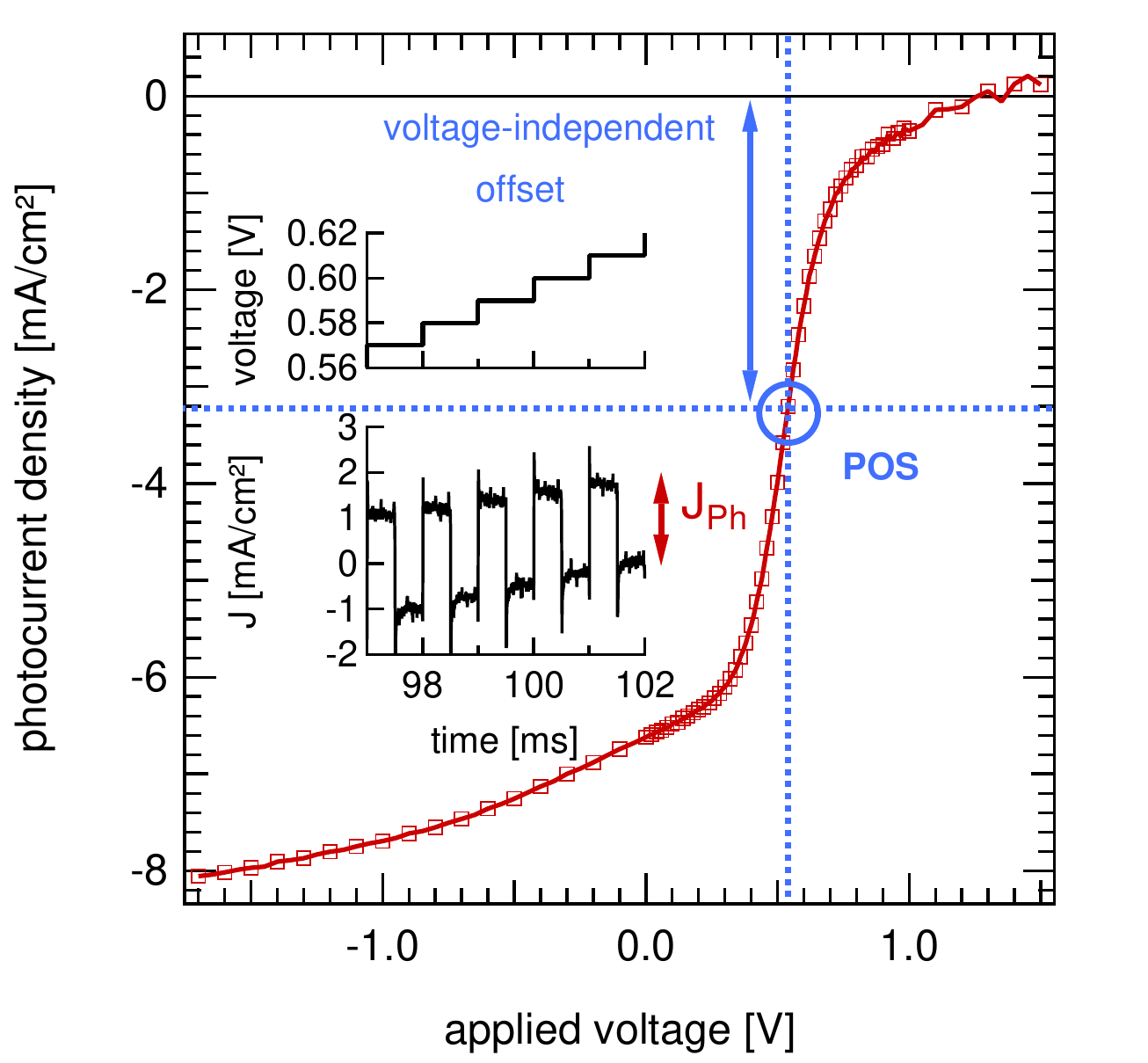}
	\caption{(Color online) Photocurrent density \jph vs. applied voltage of a typical P3HT:PCBM solar cells with Ca/Al cathode under 1 sun. The point of optimal symmetry (POS) is indicated by the blue circle. The voltage-dependent, symmetric shape of \jph is shifted by a constant offset to more negative values. Note that the density of recorded data points is higher between 0 V and 1 V applied voltage. The inset shows the current flowing under pulsed illumination, with an increase in the applied voltage every millisecond.}
	\label{fig:legend}
\end{figure}

In this paper, we investigate the symmetry and voltage dependence of the photocurrent, and compare this to a model that takes field dependent polaron pair dissociation and voltage dependent charge extraction into account. The photocurrent \jph, defined as the difference of illuminated and dark current, $J_\text{light}-J_\text{dark}$, can experimentally be accessed using pulsed illumination (inset in Figure~\ref{fig:legend}). This is necessary to avoid a major overestimation due to device heating.\cite{ooi2008}

We find the photocurrent of annealed P3HT:PCBM solar cells to have a point of optimal symmetry (POS). The photocurrent shows a point symmetry with respect to POS (see Figure~\ref{fig:legend}), and the corresponding voltage \vpos is in the range of 0.52--0.64 V.  While these values agree with results of Ooi et al.\cite{ooi2008}, our interpretation differs. We identify this voltage as the quasi flat band case--with flat bands in the bulk of the solar cell--which is well below the built-in voltage. Accordingly, our considerations lead to new insights into the origin of the photocurrent.

As shown in Figure~\ref{fig:legend}, the photocurrent is composed of two contributions. The first is voltage-dependent and symmetric with respect to POS. The second contribution is a voltage-independent offset. This constant offset is usually negative, thereby increasing the short-circuit current \jsc and PCE. As the magnitude of this offset is crucial for device performance, we investigated its dependence on thermal treatment, cathode material and active layer thickness in detail.

In order to describe the voltage dependence of \jph, we use a combination of Braun--Onsager \cite{braun1984,onsager1938} theory and charge extraction as calculated by Sokel and Hughes. \cite{sokel1982} This combination was proposed by Mihailetchi et al. \cite{mihailetchi2004a} and could for the first time explain the experimental photocurrent in reverse direction. However, they calculated effective voltage and field with respect to the physically ill-defined voltage $V_0$, at which $J_\text{Ph}=0$. As we will show, this is correct only in the special case where $V_0$ coincides with \vpos. In general, contact effects result in a lower value of \vpos. Consequently, we consider voltage and electric field relative to \vpos instead of $V_0$. Calculating polaron pair dissociation and charge extraction with respect to the point of symmetry now allows to describe the experimental photocurrent for solar cells with different electrode material. Using a pulsed measurement technique also makes the photocurrent in forward direction accessible, and shows that the model can describe the data in both directions.

\section{Experimental}

All samples investigated were spin-coated from solution of a 1:1 weight ratio of P3HT and PCBM in chlorobenzene, onto indium tin oxide/glass substrates coated with \PEDOT:\PSS. After an annealing step at 140 \degree, metal contacts were applied by thermal evaporation; either Ca (3 nm) followed by Al (110 nm) or only Ag (120 nm). All fabrication and characterization took place in a nitrogen atmosphere glovebox. The cells with Ca/Al cathode had fill-factors (FF) of about 60 \% and PCE of 2--3 \%, as determined with a Xe-lamp, adjusted to simulate standard testing conditions (STC).\cite{shrotriya2006}

The photocurrent measurements were performed based on the setup proposed by Ooi et al. \cite{ooi2008}, using a white 10 W LED for pulsed illumination. The pulse duration was set to 0.5 ms at a duty cycle of 50 \%, giving the cell ample time to reach steady-state after switching the light on/off. The illumination level was equivalent to one sun, based on comparison of the short-cuircuit current (\jsc) reached with the LED to the \jsc reached under simulated STC. The active layer thickness was determined with a Veeco Dektak 150 profilometer.

For the capacitance--voltage (CV) measurements we applied a 5 kHz AC voltage with an amplitude of 40 mV to our cells and performed a DC bias sweep. It was carried out under nitrogen atmosphere and without any incident light to the device to prevent distortions due to charge carrier generation. Data acquisition was carried out with an Agilent E4980A LCR-meter in parallel RC circuit mode. The reasonability of this working mode has been proven by preceding impedance measurements that showed a clear semicircle (with a negligible offset impedance of about 10 $\Omega$, corresponding to the series resistance $R_s$) in Cole--Cole representation for our cells.
Since we have Schottky-like contacts at the metal--semiconductor interfaces of our devices, the value of $V_\text{CV}$ was determined by the intercept of a linear fit of the Mott--Schottky plot. 

All presented simulations were done solving the one dimensional differential
equation system of Poisson`s equation and the continuity equations for electron and holes in a numerical iterative approach,\cite{selberherr1984, deibel2008a} implicitly assuming a band-like transport. A 100 nm thick
bulk-heterojunction solar cell is calculated at a temperature of $T=300$ K containing electrons and holes with equal mobilities of $\mu_n = \mu_p = 10^{-4}$ cm$^2$/Vs.\cite{baumann2008} For the active material, the effective medium
approach is used, leading to effective electrical bandgap of $E_G=1.1$ eV
between the donor HOMO (highest occupied molecular orbital) and the acceptor
LUMO (lowest unoccupied molecular orbital) level and an
effective dielectric constant of $\epsilon_r = 3.4$.\cite{baumann2008} The
effective charge carrier densities of LUMO and HOMO---describing the spatial
charge carrier densities of electrons $n(x)$ and holes $p(x)$ by
Fermi-Dirac statistics---were set to $N_c = N_v = 10^{20}$ cm$^{-3}$.
The interfaces towards the electrodes are assumed to possess injection
barriers of 0.1 eV without any surface recombination.\cite{rauh2010}

Illumination is taken into account by a homogenous charge carrier generation rate $G$ over the whole device. This simplification avoids band bending due to local differences of the charge carrier densities caused by optical interference inside the sample. We use Langevin theory to describe bimolecular charge carrier recombination of free polarons to the ground state.\cite{langevin1903} The combination of charge carrier generation and recombination leads to the following net generation rate:

\begin{equation}
	U(x)= G - \frac{q (\mu_n + \mu_p )}{\epsilon_0 \epsilon_r} \left( np - n_i^2 \right)
\end{equation}

with the elementary charge $q$, the dielectric constant $\epsilon_0$, the
Boltzmann constant $k_B$ and the intrinsic charge carrier density
$n_i=\sqrt{N_c N_v}\exp \left(q E_G / 2 k_B T \right)$.

The result of this system are spatially resolved values for the electrical
potential and electron and hole densities in form of quasi-Fermi potentials. In order to create a band structure, the potentials have to be transformed into an energy level. Since experimental values of device work functions are not available, the electron conducting contact (here aluminum) is set to a typical value of -4.25 eV,\cite{sze1981} which determines the absolute value of all other potentials.

\begin{figure}[tb]
	\centering
	\includegraphics[height=7.5cm]{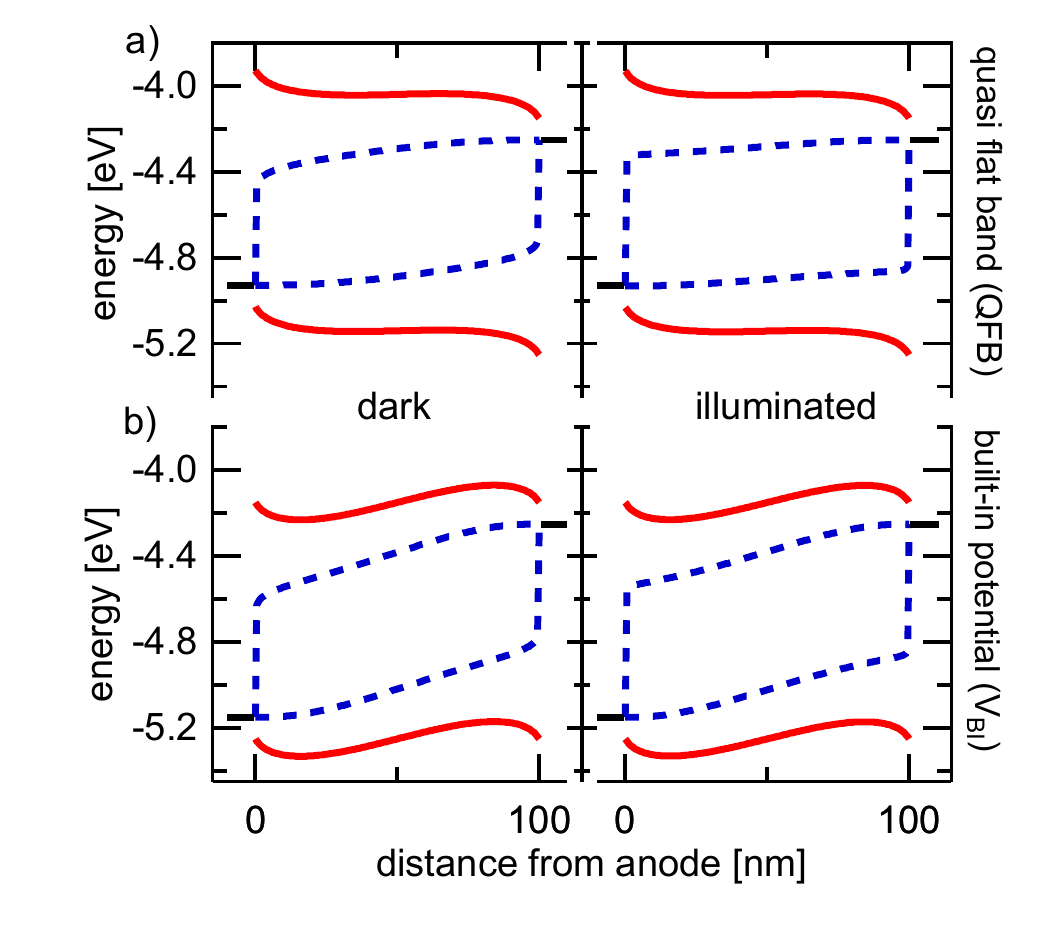}
	\caption{(Color online) Energy structures of a simulated organic solar cell in the dark (left) and under illumination of 1 sun (right). The simulated HOMO and LUMO levels (solid lines) and quasi Fermi-levels for electrons and holes (dashed lines) of a 100 nm thick cell are shown, with the ITO contact on the left and the metal contact on the right side. a) At \vpos (QFB) the bands are flat in the bulk, while the electric field is finite in the contact regions. b) At an applied voltage equal to the built-in potential \vbi the vacuum levels at the position of the electrodes match, but the internal electric field is nonzero over the whole device.}
	\label{fig:simbands}
\end{figure}

In Figure \ref{fig:simbands} the resulting band structures are presented in the dark and under illumination for two important bias voltages. First, a voltage resulting in flat bands---and therefore zero electric field $E$, being the spatial derivative of the potential---in the middle of the sample, and second the built-in voltage for comparison.

\section{Results and Discussion}

\subsection{Origin of \vpos}

The photocurrent shows a high symmetry with respect to POS, with a corresponding voltage \vpos in the range of 0.52--0.64 V for annealed solar cells with Ca/Al cathode under 1 sun. This agrees with findings of Ooi et al.,\cite{ooi2008} who reported a \vpos of 0.58--0.60 V for cells with Al cathode. 

To understand this symmetry in \jph, we look at the energy bands in the solar cell. Due to the band bending at the contacts, a case of flat bands in the whole device does not exist. However, our macroscopic simulation shows flat band conditions with zero electric field in the bulk of the cell at an applied voltage of 0.66 V (Figure~\ref{fig:simbands}a). We call this the quasi flat band (QFB) case, which is well below the built-in voltage \vbi at 0.90 V (Figure~\ref{fig:simbands}b).

The reason why the bias must be reduced from \vbi in order to achieve flat bands in the bulk is the band bending at the electrodes, which is a consequence of the boundary conditions for the quasi-Fermi levels at the interfaces. Ideally, the electrodes inject charge carriers of one type into the organic semiconductor and extract the other type. This generates a strong diffusive current in the contact region which is, in thermal equilibrium, accompanied by a high electric field in the opposite direction, especially next to the electrodes. As shown in Figure~\ref{fig:simbands}, the resulting band bending is almost independent of illumination. Only the quasi-Fermi levels are affected, indicating a higher charge carrier density under illumination.

This reduced voltage was published by Kemerink et al., but incorrectly identified as $V_0$.\cite{kemerink2006} At QFB, the electric field is zero in the bulk of the solar cell and finite in vicinity of the contacts, while at \vbi the field is finite at every position in the cell.

\begin{figure}[tb]
    \includegraphics[height=7cm]{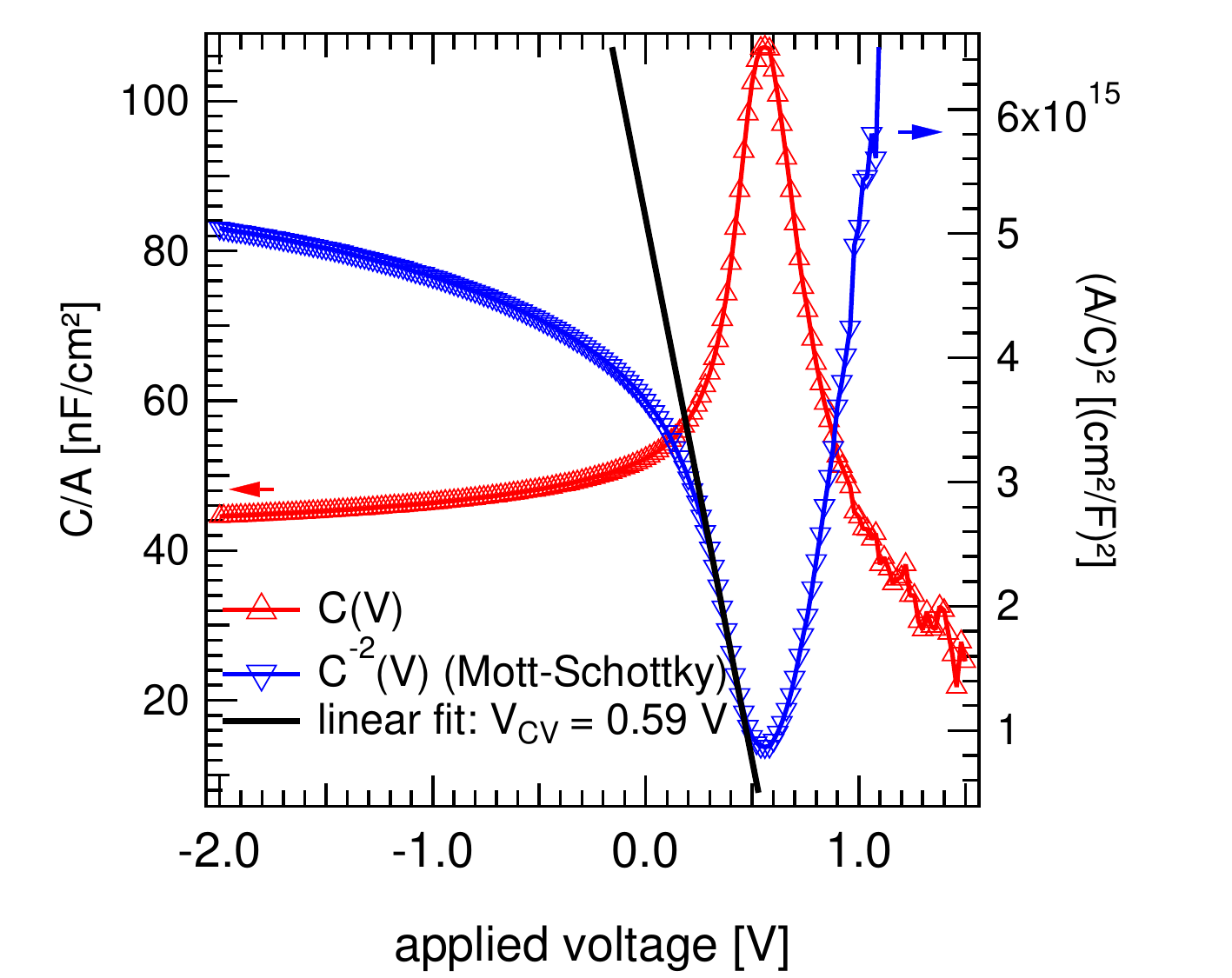}
    \caption{(Color online) Typical result for a capacitance--voltage measurement of an annealed P3HT:PCBM solar cell with Ca/Al cathode in the dark. Normalized capacitance $C/A$ (left axis) and Mott-Schottky plot (right axis). A linear fit to the Mott-Schottky plot results in a value of 0.59 V for the quasi flat band voltage.}
    \label{fig:cv}
\end{figure}

Capacitance--voltage measurements---conducted in the dark---further support the interpretation of \vpos being the quasi flat band case. The capacitance $C=\delta Q/\delta V=\delta Q/\delta (Ed)$ (where $Q$ is the electric charge) depends on the differential variation of the electric field $\delta E$ in the bulk, which approaches zero in this case whereby the capacitance approaches infinity. An extrapolation of $C^{-2}$ to zero (linear fit to the Mott--Schottky plot, see Figure~\ref{fig:cv}) yields \vcv, which therefore corresponds to the quasi flat band voltage. This is a common technique to determine the flat band voltage in inorganic semicondutors, \cite{blood} but has already been applied to organic devices.\cite{yakuphanoglu2008, bisquert2008} We found \vcv to be 0.5--0.6 V, which agrees with the observed values for \vpos.

The quasi flat band case is distinct and can explain the symmetry of the photocurrent. Energy bands with zero slope present a special, symmetric situation, as a relative bias in either direction will cause the photogenerated carriers in the bulk to flow equally strong, but in different directions.
In conclusion of this, we propose \vpos to correspond to the case of flat bands in the bulk of the active layer, with finite electric field in vicinity of the contacts, caused by band bending due to injection barriers. Hence, the electric field in the bulk will be proportional to an effective voltage $|V-V_\text{POS}|$. In contrast to Ooi et al., \cite{ooi2008} we do not assume \vpos to equal the built-in voltage \vbi, but to be considerably smaller.

\subsection{Voltage-independent offset of \jph}

\begin{figure}[tb]
\begin{minipage}[t]{.48\linewidth}
\includegraphics[height=6cm]{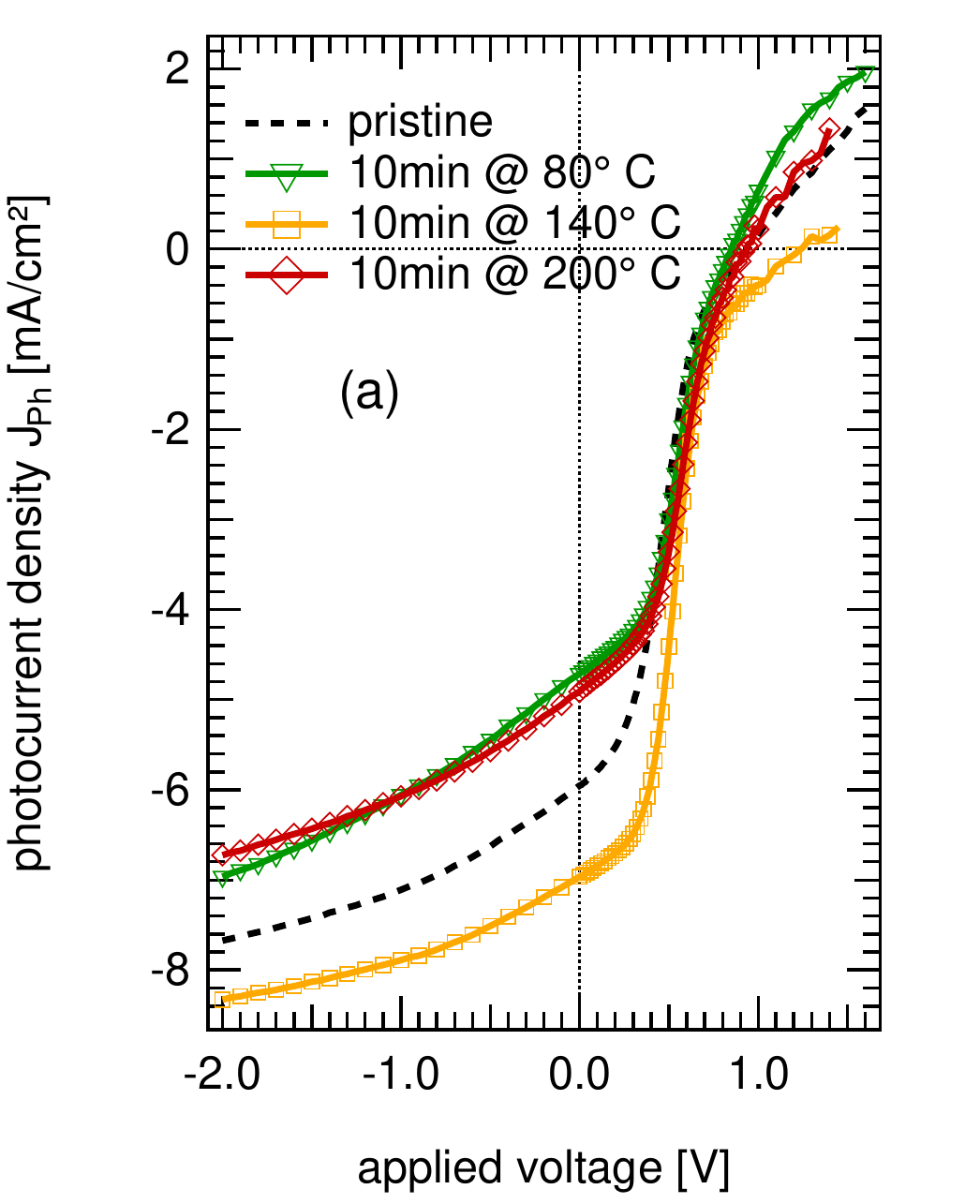} 
\end{minipage}
\begin{minipage}[t]{.48\linewidth}
\includegraphics[height=6cm]{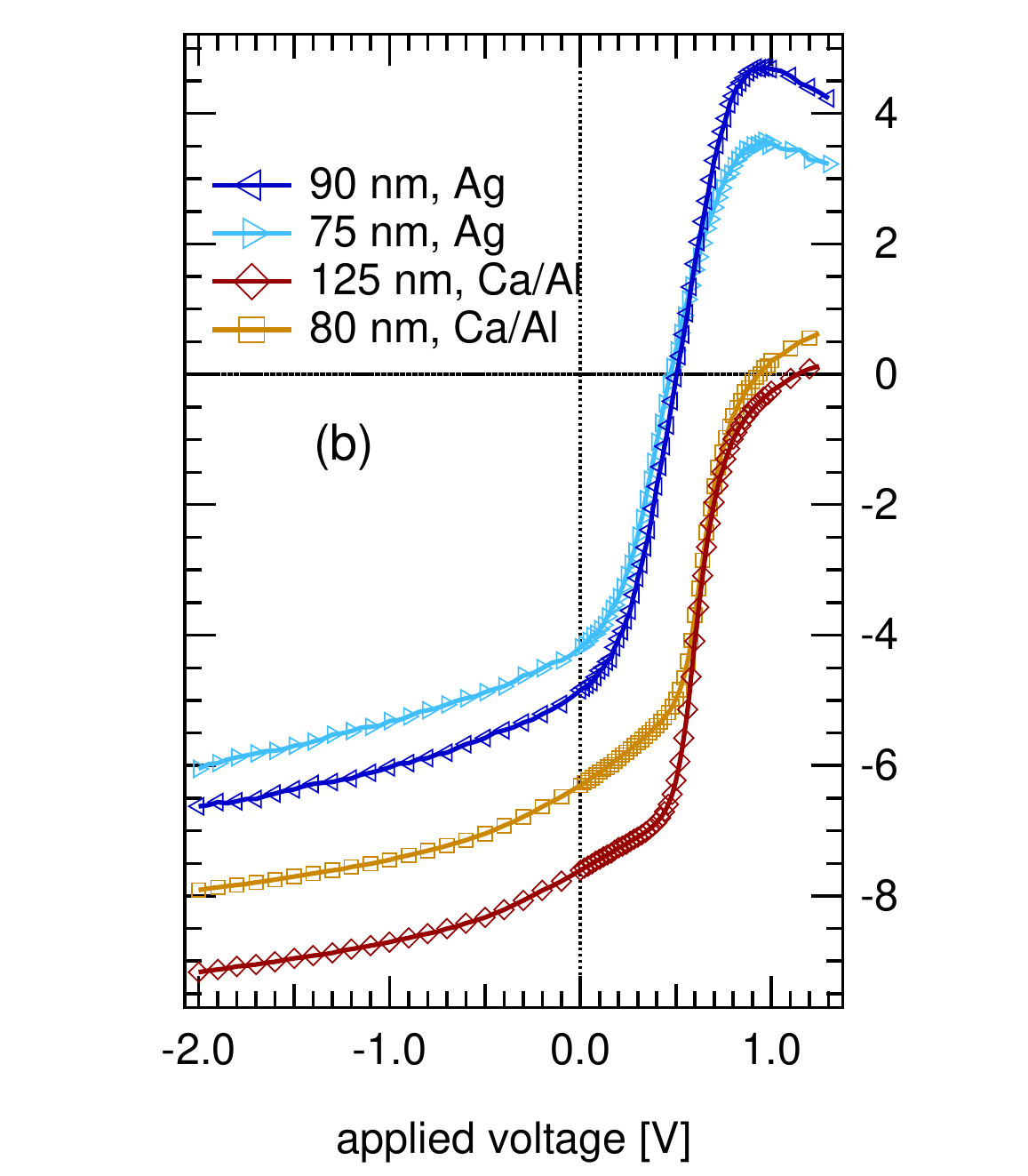} 
\end{minipage}
\caption{(Color online) Photocurrent density vs. applied voltage. (a) Influence of thermal treatment for solar cells with Ca/Al cathode. (b) Cells with Ca/Al and Ag cathode, in each case for two different active layer thicknesses. The magnitude of the voltage-independent offset can be influenced by thermal treatment and choice of cathode material.}
\label{fig:offsets}
\end{figure}

As mentioned above, the photocurrent is composed of two contributions. The first is voltage-dependent and symmetric with respect to POS. The second contribution is a voltage-independent offset, which is crucial for device performance and can be influenced by processing parameters. Figure~\ref{fig:offsets} shows the influence of cathode material and annealing conditions on the photocurrent. While the shape of \jph remains similar, the offset is critically dependent on both cathode material and thermal treatment. With respect to the pristine sample, the offset decreases after thermal treatment at 80 or 200 \degree, but increases upon treatment at 140 \degree (Figure~\ref{fig:offsets}a). 
Using Ag as cathode material yields a much smaller offset than Ca/Al, while the shape of the photocurrent curves is very similar (Figure~\ref{fig:offsets}b). \vpos is also smaller for cells with Ag cathode, with values of about 0.45 V, which is a direct consequence of the higher injection barriers.
Active layer thickness $d$ influences the magnitude of the offset as well. The highest negative offset, up to 4.1 $\mathrm{mA~cm}^{-1}$, was observed for solar cells with Ca/Al cathode, active layer thickness of 120 nm and thermal annealing at 140 \degree. These cells had the highest \jsc and PCE.

As this voltage-independent offset can be influenced by the choice of cathode material without changing the shape of the photocurrent significantly, we assume the contact regions to be responsible for the offset. The voltage dependent part of \jph then corresponds to the photocurrent generated in the bulk of the solar cell.

Ooi et al. explained this offset with self-selective electrodes,\cite{ooi2008} resulting in a constant diffusion current \mbox{$J_\text{Ph, offset}=-(eDn_\text{ph}/d)$}, where $e$ is the elementary charge, $D$ the diffusivity and $n_{ph}$ the concentration of photogenerated charge carriers at the selective electrode. In contrast to this predicted $d^{-1}$ dependence, we observed a maximum (negative) offset for a thickness $d$ of about 120 nm, with smaller offsets for thinner and thicker devices (not shown).

As an alternative explanation for this constant offset, we propose the band bending in vicinity of the contacts, which---independently of applied bias---has only one direction and results in an electric field high enough for efficient polaron pair dissociation (see Figure~\ref{fig:simbands}). The contact regions would then give a constant contribution to \jph, which depends on the degree of band bending---and therefore the height of the injections barriers---while the bulk region contributes to \jph as a function of applied voltage and causes the symmetry.

The exact nature of this offset and its dependence on the cathode material has to be further investigated. A highly negative offset is crucial for device performance, and should be optimized, e.g. by choice of cathode material.

\subsection{Voltage vs. field dependence}

\begin{figure}[tb]
\begin{minipage}[t]{.5\linewidth}
\includegraphics[height=6cm]{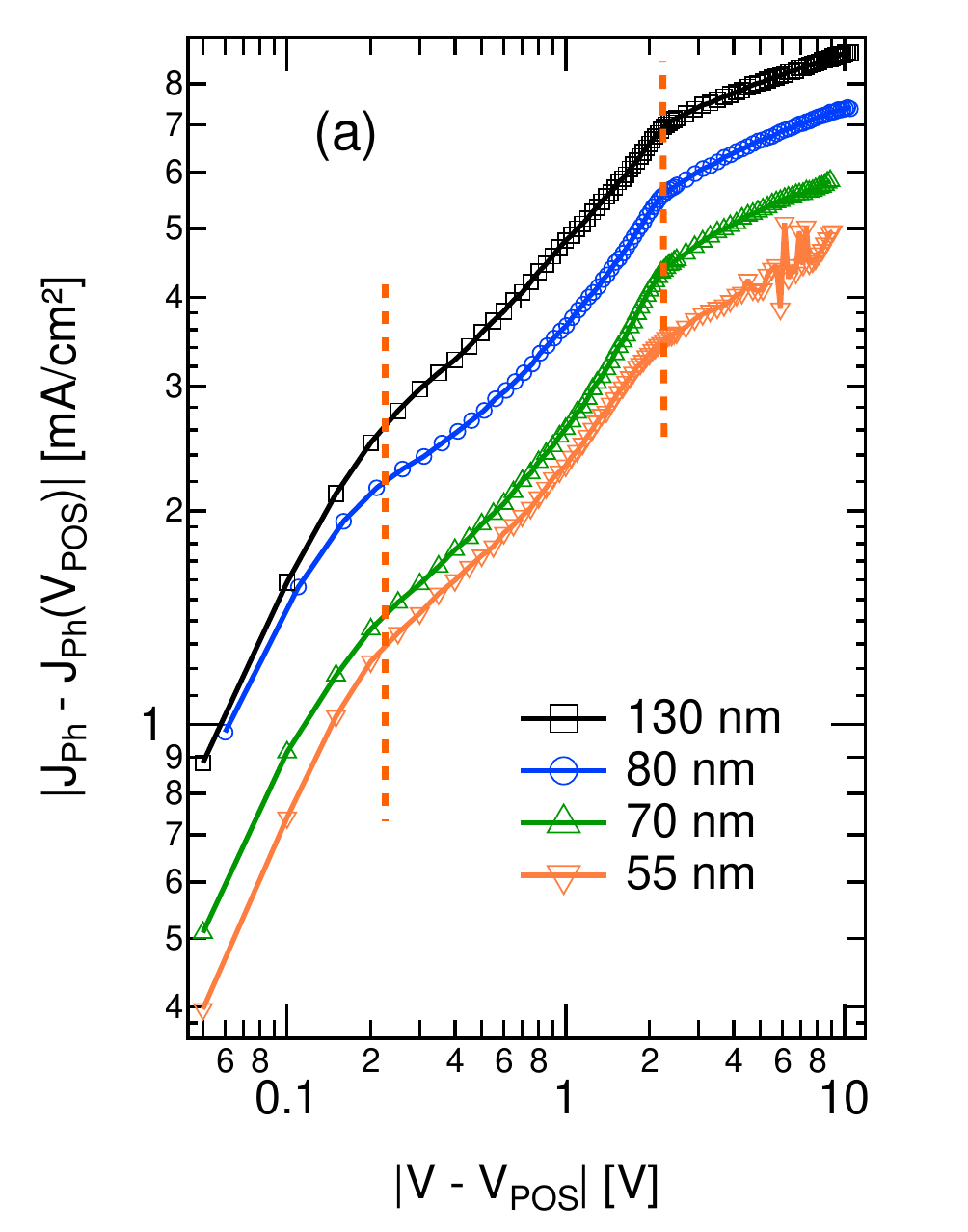} 
\end{minipage}
\begin{minipage}[t]{.44\linewidth}
\includegraphics[height=6cm]{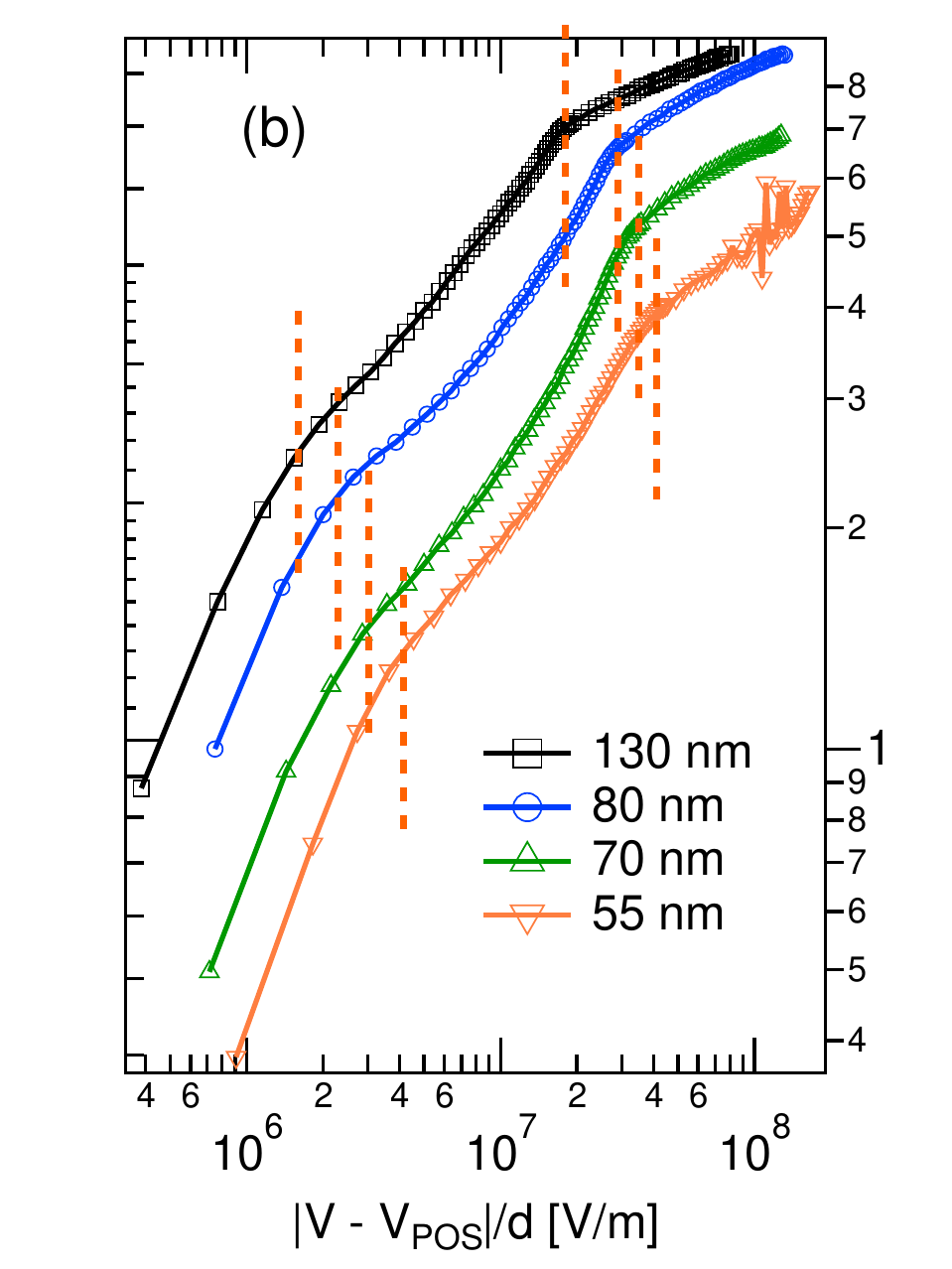} 
\end{minipage}
	\caption{(Color online) Plot of the photocurrent in reverse direction (with the origin shifted to POS) against effective voltage (a) and calculated field (b) for cells with different active layer thicknesses. While the characteristic points, indicated by the red dashed lines, lie on top of each other in the left graph, they disperse in the right graph.}
	\label{fig:thickness}
\end{figure}

The active layer thickness $d$ has only little influence on the voltage dependence of the observed photocurrent (Figure~\ref{fig:thickness}(a)). Of course the absolute value of \jph is higher for thicker cells, but the shape of $J_\text{Ph}(V)$ is very similar for cells with thicknesses between 55 and 130 nm. This is surprising, since e.g. polaron pair dissociation and thereby \jph are supposed to be dependent on electric field $E$.\cite{deibel2009} In the simplest approximation, assuming a constant slope in energy bands over the whole extension of the device, the effective field is the fraction of effective voltage and active layer thickness:

\begin{equation}
E=|V-V_\text{POS}|/d
\label{eq:field}
\end{equation}

Characteristic points of the photocurrent, indicated by red dashed lines in Figure~\ref{fig:thickness}, lie on top of each other when plotted against effective voltage $|V-V_\text{POS}|$. When plotted against the calculated field $E$ (Equation~(\ref{eq:field})), the curves of \jph disperse (Figure~\ref{fig:thickness}(b)).
This indicates that the approximation for the internal field in Equation~(\ref{eq:field}) is not realistic. The reason for this is the voltage drop at the contacts (Figure~\ref{fig:simbands}), which greatly reduces the electric field in the bulk of the cell. In addition the photocurrent is not only governed by field dependent polaron pair dissociation, but also depends on the extraction of charge carriers. As described below, the active layer thickness does not influence this extraction mechanism.

\subsection{Polaron pair dissociation and charge extraction}

To describe the voltage dependence of the photocurrent, we propose a model that takes field-dependent polaron pair (PP) dissociation and voltage dependent polaron extraction by drift and diffusion into account.
A well-known model to describe field-dependent PP dissociation was presented by Braun in 1984.\cite{braun1984} Using the field dependence for ion-pair dissociation by Onsager and Langevin recombination for polarons,\cite{onsager1938,langevin1903} Braun derives the following PP dissociation probability:

\begin{equation}
  \centering
	P_\text{Braun}(E)=\frac{k_d(E)}{k_d(E)+k_\text{f}}=\frac{\kappa_d(E)}{\kappa_d(E)+(\mu\tau_\text{f})^{-1}}~,
	\label{eq:Braun}
\end{equation}

where $k_d(E)=\kappa_d(E)\mu$ is the PP dissociation rate. It can be calculated by detailed balance from the recombination rate and the Coulomb binding energy, which depends on the PP radius \rpp. $k_\text{f}$ is the decay rate of a PP to the ground state. Since $\mu$ and $\tau_\text{f}=k_\text{f}^{-1}$ do not influence $P_\text{Braun}(E)$ independently, we use their product \mutau as a single parameter. \cite{deibel2009}

Upon successful separation, the polarons need to travel to the electrodes to be extracted. This part can be described with a term introduced by Sokel and Hughes.\cite{sokel1982} They solved the problem of photoconductivity in insulators analytically, neglecting dark current, trapping and recombination. The negligence of recombination is a resonable approximation.\cite{pivrikas2005,deibel2008b,deibel2009prb} The result depends on temperature $T$ and applied voltage $V$:

\begin{equation}
  \centering
	J_\text{Ph}=J_\text{Ph,max}\left[\frac{\exp(qV/kT+1)}{\exp(qV/kT-1)}-\frac{2kT}{qV}\right]
	\label{eq:SH}
\end{equation}

Using an effective voltage $V_\text{eff}=|V-V_\text{POS}|$ and a calculated field $E=|V_\text{eff}|/d$, and adding the constant offset \offset, the overall photocurrent can be written as:

\begin{equation}
  \centering
	J_\text{Ph}=J_\text{Ph,max}\left[\coth\left(\frac{eV_\text{eff}}{2kT}\right)-\frac{2kT}{eV_\text{eff}}\right] P_\text{Braun}(E) + J_\text{Ph}(V_\text{POS})
	\label{eq:jph_model}
\end{equation}

This combination of Braun-model for polaron pair dissociation and the Sokel--Hughes term for charge extraction was first proposed by Mihailetchi et al.,\cite{mihailetchi2004a} but with the assumption that $V_\text{eff}=|V-V_0|$. This corresponds to the special case of a vanishing offset, where \vpos equals $V_0$. Also, they used a Gaussian distribution of polaron pair radii \rpp to calculate the dissociation efficiency in the Braun model. While this assumption seems reasonable, a fixed value for \rpp could describe our data better than a distributed one.

\begin{figure}[tb]\centering
	\includegraphics[width=8cm]{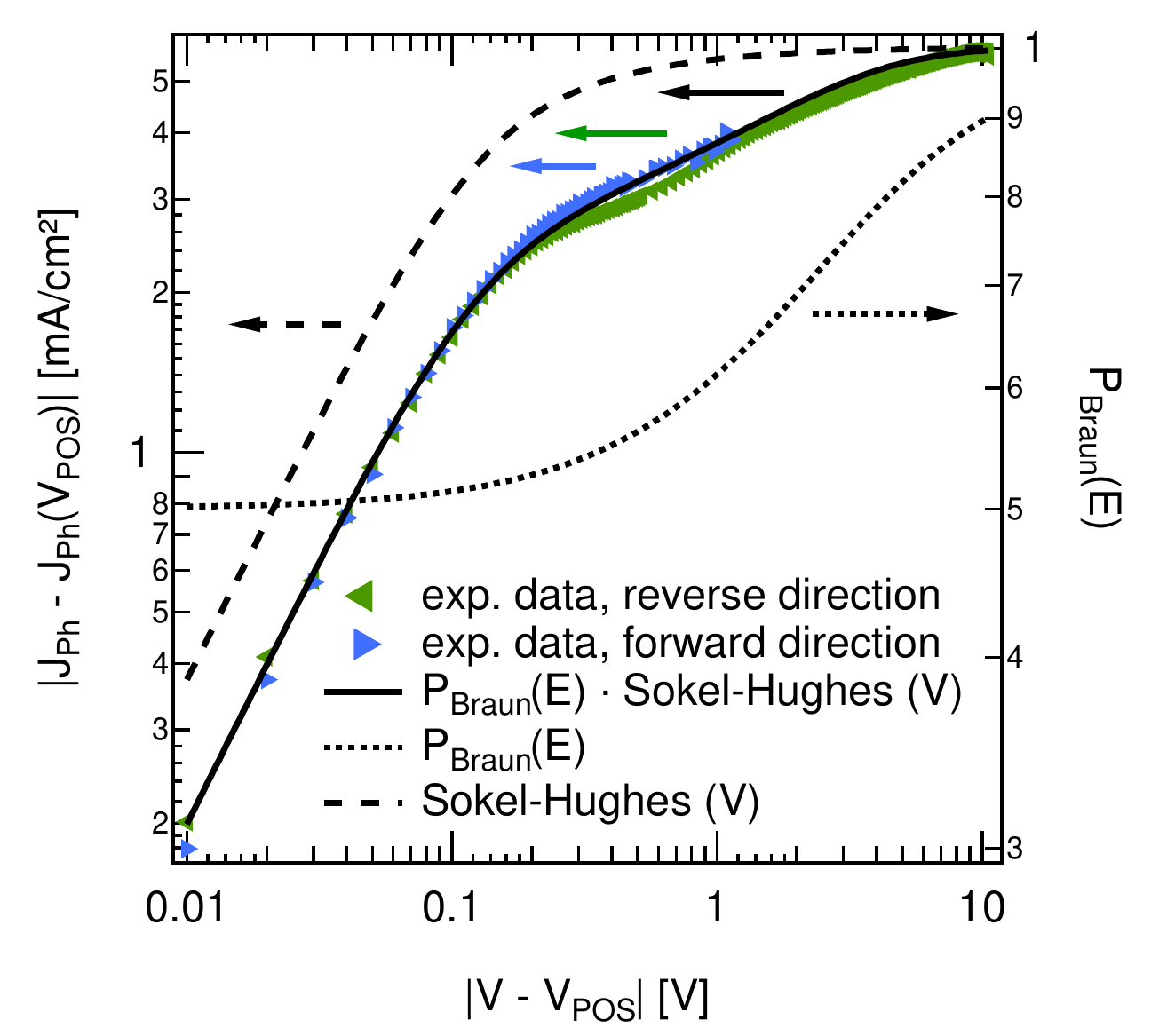}
	\caption{(Color online) Comparison of experimental photocurrent (cell with Ca/Al cathode under 1 sun) and a model based on Braun--Onsager theory and extraction. Field dependent part of \jph in forward and reverse direction (left axis) and PP dissociation probability (right axis) vs. effective voltage. The black lines show the PP dissociation after Braun (dotted), extraction after Sokel \& Hughes (dashed) and their combined product (solid).}
	\label{fig:sym}
\end{figure}

\begin{figure}[tb]\centering
	\includegraphics[width=7.5cm]{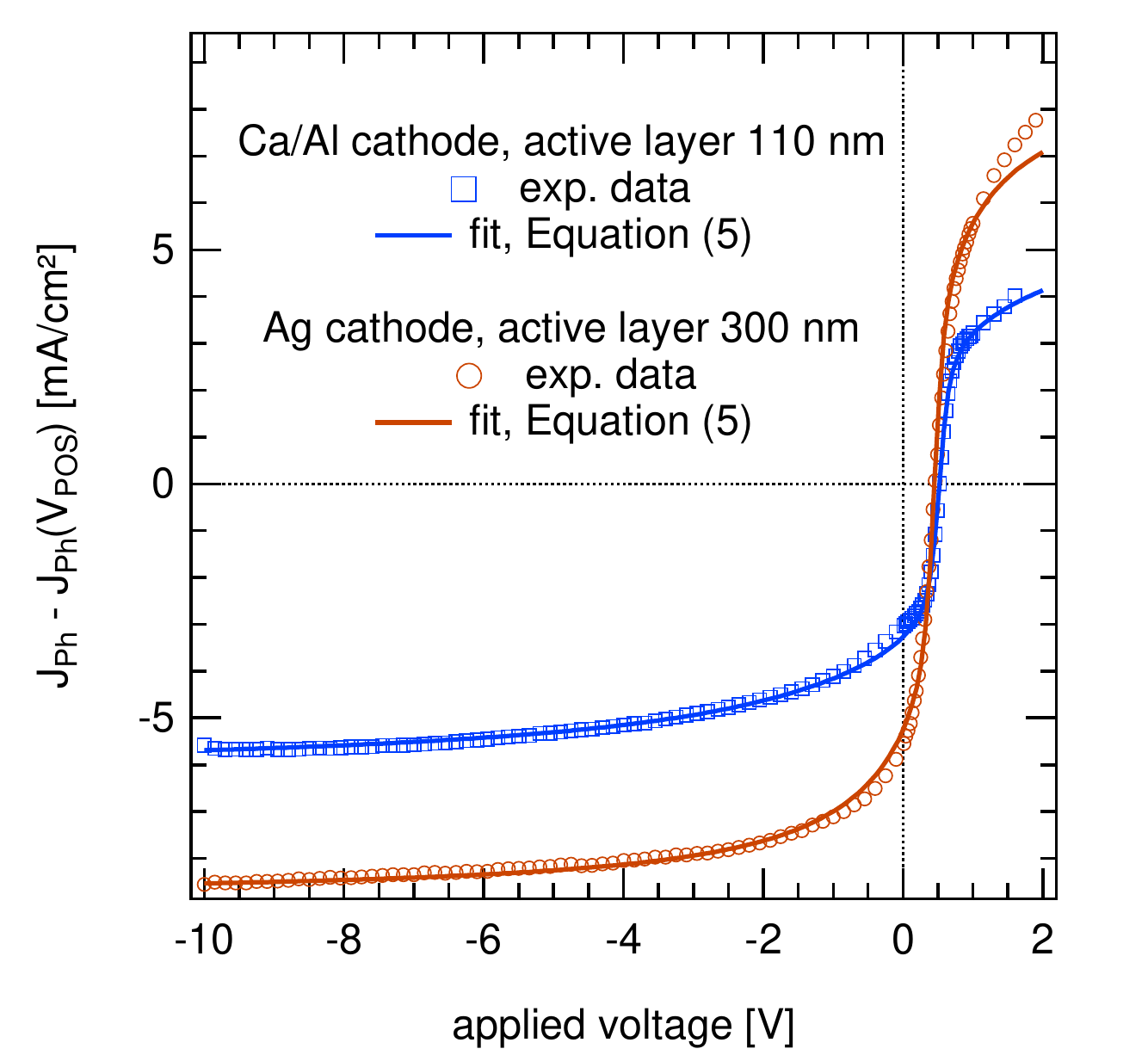}
	\caption{(Color online) Effective photocurrent $J_\text{Ph}-J_\text{Ph}(V_\text{POS})$ vs. applied voltage for two solar cells with Ca/Al and Ag cathode (symbols). For both cells, the experimental data (symbols) agrees well with the model, Equation~(\ref{eq:jph_model}) (solid lines).}
	\label{fig:linfit}
\end{figure}

As shown in Figure~\ref{fig:sym}, the product of PP dissociation after Braun and extraction after Sokel--Hughes (Equation~(\ref{eq:jph_model})) describes the experimental photocurrent well in both forward and reverse directions. 
A set of parameters in a narrow range could be employed to describe the experimental photocurrent of solar cells with different active layer thickness and cathode material over the whole measured range (Figure~\ref{fig:linfit}).

Using Equation~(\ref{eq:field}) to calculate $E$, effective values of about 400 nm for $d$ were used in the model, almost independent of measured device thickness and clearly exceeding it. This is a consequence of the voltage drops at the contacts (Figure~\ref{fig:simbands}). Low injection barriers result in a high band bending in vicinity of the electrodes and reduced electric field in the bulk of the cell, corresponding to a seemingly greater $d$. The approximation $E=V_\text{eff}/d$ is therefore flawed.

The saturated, voltage-dependent part of the photocurrent $J_\text{Ph,max}$ is typically 5--8 $\mathrm{mA~cm}^{-2}$, with higher values for thicker cells. The dielectric constant $\epsilon_r$ was set to 3.5 and temperature $T$ to 310 K (slightly above room temperature to account for heating under illumination). The \mutau product of spatially averaged mobility $\mu$ and PP lifetime (with respect to the ground state) $\tau_f=1/k_f$ was set to 1--3$\cdot 10^{-14}\mathrm{m^2~V^{-1}}$ and the PP radius \rpp to 1.8--2.0 nm. 

Using this parameter set, the dissociation efficiency of polaron pairs in the bulk can be calculated to be 40--60 \% at zero field. However, extraction is the limiting factor at low effective voltages. Only at higher effective voltages, the field dependence of polaron pair dissociation determines the shape of the photocurrent.

\section{Conclusions}

By use of a pulsed measurement technique, we find a point of symmetry for the photocurrent at 0.52--0.64 V in P3HT:PCBM organic solar cells, which agrees with results from capacitance--voltage measurements. Based on our macroscopic simulation, we identify this voltage as the point of flat bands in the bulk of the solar cell, the quasi flat band case, well below the built-in potential. We find a voltage-independent offset of the photocurrent, which is crucial for high \jsc, which depends both on thermal annealing conditions and cathode material. We also propose a possible origin of this offset with the band-bending close to the contacts. Active layer thickness has an influence on this offset, but surprisingly little impact on the shape of \jph. 

A model including polaron pair dissociation, based on Braun--Onsager theory, and extraction, based on the work by Sokel \& Hughes, can describe our data for different solar cells with a narrow range of parameters. The electric field in the bulk of the cell is greatly reduced by the injection barriers. At zero electric field, polaron pair dissociation efficiency is 40--60 \%.

Since the maximum power point is close to \vpos, where the effective voltage is zero, the photocurrent under operating conditions is largely determined by the voltage independent offset. For best device performance it is necessary to maximize this constant contribution from the contact regions and simultaneously enhance polaron pair dissociation and polaron extraction in the bulk.

\begin{acknowledgments}

The authors thank the Bundesministerium f\"{u}r Bildung und Forschung for financial support in the framework of the MOPS project (contract no.\ 13N9867). V.D.'s work at the ZAE Bayern is financed by the Bavarian Ministry of Economic Affairs, Infrastructure, Transport and Technology. 

\end{acknowledgments}


\begin{thebibliography}{10}%
\makeatletter
\providecommand \@ifxundefined [1]{%
 \ifx #1\undefined \expandafter \@firstoftwo
 \else \expandafter \@secondoftwo
\fi
}%
\providecommand \@ifnum [1]{%
 \ifnum #1\expandafter \@firstoftwo
 \else \expandafter \@secondoftwo
\fi
}%
\providecommand \enquote [1]{``#1''}%
\providecommand \bibnamefont  [1]{#1}%
\providecommand \bibfnamefont [1]{#1}%
\providecommand \citenamefont [1]{#1}%
\providecommand\href[0]{\@sanitize\@href}%
\providecommand\@href[1]{\endgroup\@@startlink{#1}\endgroup\@@href}%
\providecommand\@@href[1]{#1\@@endlink}%
\providecommand \@sanitize [0]{\begingroup\catcode`\&12\catcode`\#12\relax}%
\@ifxundefined \pdfoutput {\@firstoftwo}{%
 \@ifnum{\z@=\pdfoutput}{\@firstoftwo}{\@secondoftwo}%
}{%
 \providecommand\@@startlink[1]{\leavevmode}%
 \providecommand\@@endlink[0]{}%
}{%
 \providecommand\@@startlink[1]{%
  \leavevmode
  \pdfstartlink
   attr{/Border[0 0 1 ]/H/I/C[0 1 1]}%
   user{/Subtype/Link/A<</Type/Action/S/URI/URI(#1)>>}%
  \relax
 }%
 \providecommand\@@endlink[0]{\pdfendlink}%
}%
\providecommand \url  [0]{\begingroup\@sanitize \@url }%
\providecommand \@url [1]{\endgroup\@href {#1}{\urlprefix}}%
\providecommand \urlprefix [0]{URL }%
\providecommand \Eprint[0]{\href }%
\@ifxundefined \urlstyle {%
  \providecommand \doi [1]{doi:\discretionary{}{}{}#1}%
}{%
  \providecommand \doi [0]{doi:\discretionary{}{}{}\begingroup
  \urlstyle{rm}\Url }%
}%
\providecommand \doibase [0]{http://dx.doi.org/}%
\providecommand \Doi[1]{\href{\doibase#1}}%
\providecommand \bibAnnote [3]{%
  \BibitemShut{#1}%
  \begin{quotation}\noindent
    \textsc{Key:}\ #2\\\textsc{Annotation:}\ #3%
  \end{quotation}%
}%
\providecommand \bibAnnoteFile [2]{%
  \IfFileExists{#2}{\bibAnnote {#1} {#2} {\input{#2}}}{}%
}%
\providecommand \typeout [0]{\immediate \write \m@ne }%
\providecommand \selectlanguage [0]{\@gobble}%
\providecommand \bibinfo [0]{\@secondoftwo}%
\providecommand \bibfield [0]{\@secondoftwo}%
\providecommand \translation [1]{[#1]}%
\providecommand \BibitemOpen[0]{}%
\providecommand \bibitemStop [0]{}%
\providecommand \bibitemNoStop [0]{.\EOS\space}%
\providecommand \EOS [0]{\spacefactor3000\relax}%
\providecommand \BibitemShut [1]{\csname bibitem#1\endcsname}%
%</preamble>
\bibitem{Park2009}%
  \BibitemOpen
  \bibfield{author}{%
  \bibinfo {author} {\bibfnamefont{S.~H.}\ \bibnamefont{Park}}, \bibinfo
  {author} {\bibfnamefont{A.}~\bibnamefont{Roy}}, \bibinfo {author}
  {\bibfnamefont{S.}~\bibnamefont{Beaupre}}, \bibinfo {author}
  {\bibfnamefont{S.}~\bibnamefont{Cho}}, \bibinfo {author}
  {\bibfnamefont{N.}~\bibnamefont{Coates}}, \bibinfo {author}
  {\bibfnamefont{J.~S.}\ \bibnamefont{Moon}}, \bibinfo {author}
  {\bibfnamefont{D.}~\bibnamefont{Moses}}, \bibinfo {author}
  {\bibfnamefont{M.}~\bibnamefont{Leclerc}}, \bibinfo {author}
  {\bibfnamefont{K.}~\bibnamefont{Lee}},\ and\ \bibinfo {author}
  {\bibfnamefont{A.~J.}\ \bibnamefont{Heeger}},\ }%
  \bibfield{journal}{%
  \bibinfo {journal} {Nature Photonics}\ }%
  \textbf{\bibinfo {volume} {3}},\ \bibinfo {pages} {297} (\bibinfo {year}
  {2009})%
  \bibAnnoteFile{NoStop}{Park2009}%
\bibitem{schilinsky2004}%
  \BibitemOpen
  \bibfield{author}{%
  \bibinfo {author} {\bibfnamefont{P.}~\bibnamefont{Schilinsky}}, \bibinfo
  {author} {\bibfnamefont{C.}~\bibnamefont{Waldauf}}, \bibinfo {author}
  {\bibfnamefont{J.}~\bibnamefont{Hauch}},\ and\ \bibinfo {author}
  {\bibfnamefont{C.~J.}\ \bibnamefont{Brabec}},\ }%
  \bibfield{journal}{%
  \bibinfo {journal} {J. Appl. Phys.}\ }%
  \textbf{\bibinfo {volume} {95}},\ \bibinfo {pages} {2816} (\bibinfo {year}
  {2004})%
  \bibAnnoteFile{NoStop}{schilinsky2004}%
\bibitem{deibel2009c}%
  \BibitemOpen
  \bibfield{author}{%
  \bibinfo {author} {\bibfnamefont{C.}~\bibnamefont{Deibel}},\ }%
  \bibfield{journal}{%
  \bibinfo {journal} {Phys. Stat. Sol. A}\ }%
  \textbf{\bibinfo {volume} {206}},\ \bibinfo {pages} {2731} (\bibinfo {year}
  {2009})%
  \bibAnnoteFile{NoStop}{deibel2009c}%
\bibitem{sariciftci1992}%
  \BibitemOpen
  \bibfield{author}{%
  \bibinfo {author} {\bibfnamefont{N.~S.}\ \bibnamefont{Sariciftci}}, \bibinfo
  {author} {\bibfnamefont{L.}~\bibnamefont{Smilowitz}}, \bibinfo {author}
  {\bibfnamefont{A.~J.}\ \bibnamefont{Heeger}},\ and\ \bibinfo {author}
  {\bibfnamefont{F.}~\bibnamefont{Wudl}},\ }%
  \bibfield{journal}{%
  \bibinfo {journal} {Science}\ }%
  \textbf{\bibinfo {volume} {258}},\ \bibinfo {pages} {1474} (\bibinfo {year}
  {1992})%
  \bibAnnoteFile{NoStop}{sariciftci1992}%
\bibitem{ooi2008}%
  \BibitemOpen
  \bibfield{author}{%
  \bibinfo {author} {\bibfnamefont{Z.~E.}\ \bibnamefont{Ooi}}, \bibinfo
  {author} {\bibfnamefont{R.}~\bibnamefont{Jin}}, \bibinfo {author}
  {\bibfnamefont{J.}~\bibnamefont{Huang}}, \bibinfo {author}
  {\bibfnamefont{Y.~F.}\ \bibnamefont{Loo}}, \bibinfo {author}
  {\bibfnamefont{A.}~\bibnamefont{Sellinger}},\ and\ \bibinfo {author}
  {\bibfnamefont{J.~C.}\ \bibnamefont{deMello}},\ }%
  \bibfield{journal}{%
  \bibinfo {journal} {Journal of Materials Chemistry}\ }%
  \textbf{\bibinfo {volume} {18}},\ \bibinfo {pages} {1644} (\bibinfo {year}
  {2008})%
  \bibAnnoteFile{NoStop}{ooi2008}%
\bibitem{braun1984}%
  \BibitemOpen
  \bibfield{author}{%
  \bibinfo {author} {\bibfnamefont{C.~L.}\ \bibnamefont{Braun}},\ }%
  \bibfield{journal}{%
  \bibinfo {journal} {J. Chem. Phys.}\ }%
  \textbf{\bibinfo {volume} {80}},\ \bibinfo {pages} {4157} (\bibinfo {year}
  {1984})%
  \bibAnnoteFile{NoStop}{braun1984}%
\bibitem{onsager1938}%
  \BibitemOpen
  \bibfield{author}{%
  \bibinfo {author} {\bibfnamefont{L.}~\bibnamefont{Onsager}},\ }%
  \bibfield{journal}{%
  \bibinfo {journal} {Phys. Rev.}\ }%
  \textbf{\bibinfo {volume} {54}},\ \bibinfo {pages} {554} (\bibinfo {year}
  {1938})%
  \bibAnnoteFile{NoStop}{onsager1938}%
\bibitem{sokel1982}%
  \BibitemOpen
  \bibfield{author}{%
  \bibinfo {author} {\bibfnamefont{R.}~\bibnamefont{Sokel}}\ and\ \bibinfo
  {author} {\bibfnamefont{R.~C.}\ \bibnamefont{Hughes}},\ }%
  \bibfield{journal}{%
  \bibinfo {journal} {J. Appl. Phys.}\ }%
  \textbf{\bibinfo {volume} {53}},\ \bibinfo {pages} {7414} (\bibinfo {year}
  {1982})%
  \bibAnnoteFile{NoStop}{sokel1982}%
\bibitem{mihailetchi2004a}%
  \BibitemOpen
  \bibfield{author}{%
  \bibinfo {author} {\bibfnamefont{V.~D.}\ \bibnamefont{Mihailetchi}}, \bibinfo
  {author} {\bibfnamefont{L.~J.~A.}\ \bibnamefont{Koster}}, \bibinfo {author}
  {\bibfnamefont{J.~C.}\ \bibnamefont{Hummelen}},\ and\ \bibinfo {author}
  {\bibfnamefont{P.~W.~M.}\ \bibnamefont{Blom}},\ }%
  \bibfield{journal}{%
  \bibinfo {journal} {Phys. Rev. Lett.}\ }%
  \textbf{\bibinfo {volume} {93}},\ \bibinfo {pages} {216601} (\bibinfo {year}
  {2004})%
  \bibAnnoteFile{NoStop}{mihailetchi2004a}%
\bibitem{shrotriya2006}%
  \BibitemOpen
  \bibfield{author}{%
  \bibinfo {author} {\bibfnamefont{V.}~\bibnamefont{Shrotriya}}, \bibinfo
  {author} {\bibfnamefont{G.}~\bibnamefont{Li}}, \bibinfo {author}
  {\bibfnamefont{Y.}~\bibnamefont{Yao}}, \bibinfo {author}
  {\bibfnamefont{T.}~\bibnamefont{Moriarty}}, \bibinfo {author}
  {\bibfnamefont{K.}~\bibnamefont{Emery}},\ and\ \bibinfo {author}
  {\bibfnamefont{Y.}~\bibnamefont{Yang}},\ }%
  \bibfield{journal}{%
  \Doi{10.1002/adfm.200600489}{\bibinfo {journal} {Adv. Funct. Mater.}}\ }%
  \textbf{\bibinfo {volume} {16}},\ \bibinfo {pages} {2016} (\bibinfo {year}
  {2006})%
  \bibAnnoteFile{NoStop}{shrotriya2006}%
\bibitem{selberherr1984}%
  \BibitemOpen
  \bibfield{author}{%
  \bibinfo {author} {\bibfnamefont{S.}~\bibnamefont{Selberherr}},\ }%
  \emph{\bibinfo {title} {{Analysis and simulation of semiconductor devices}}}\
  (\bibinfo {publisher} {Springer},\ \bibinfo {year} {1984})%
  \bibAnnoteFile{NoStop}{selberherr1984}%
\bibitem{deibel2008a}%
  \BibitemOpen
  \bibfield{author}{%
  \bibinfo {author} {\bibfnamefont{C.}~\bibnamefont{Deibel}}, \bibinfo {author}
  {\bibfnamefont{A.}~\bibnamefont{Wagenpfahl}},\ and\ \bibinfo {author}
  {\bibfnamefont{V.}~\bibnamefont{Dyakonov}},\ }%
  \bibfield{journal}{%
  \Doi{10.1002}{\bibinfo {journal} {phys. stat. sol. (RRL) 2}}\ }%
  \textbf{\bibinfo {volume} {4}},\ \bibinfo {pages} {175} (\bibinfo {year}
  {2008})%
  \bibAnnoteFile{NoStop}{deibel2008a}%
\bibitem{baumann2008}%
  \BibitemOpen
  \bibfield{author}{%
  \bibinfo {author} {\bibfnamefont{A.}~\bibnamefont{Baumann}}, \bibinfo
  {author} {\bibfnamefont{J.}~\bibnamefont{Lorrmann}}, \bibinfo {author}
  {\bibfnamefont{C.}~\bibnamefont{Deibel}},\ and\ \bibinfo {author}
  {\bibfnamefont{V.}~\bibnamefont{Dyakonov}},\ }%
  \bibfield{journal}{%
  \bibinfo {journal} {Appl. Phys. Lett.}\ }%
  \textbf{\bibinfo {volume} {93}},\ \bibinfo {pages} {252104} (\bibinfo {year}
  {2008})%
  \bibAnnoteFile{NoStop}{baumann2008}%
\bibitem{rauh2010}%
  \BibitemOpen
  \bibfield{author}{%
  \bibinfo {author} {\bibfnamefont{D.}~\bibnamefont{Rauh}}, \bibinfo {author}
  {\bibfnamefont{C.}~\bibnamefont{Deibel}},\ and\ \bibinfo {author}
  {\bibfnamefont{V.}~\bibnamefont{Dyakonov}},\ }%
  \bibinfo {note} {unpublished}%
  \bibAnnoteFile{NoStop}{rauh2010}%
\bibitem{langevin1903}%
  \BibitemOpen
  \bibfield{author}{%
  \bibinfo {author} {\bibfnamefont{P.}~\bibnamefont{Langevin}},\ }%
  \bibfield{journal}{%
  \bibinfo {journal} {{Ann. Chim. Phys.}}\ }%
  \textbf{\bibinfo {volume} {28}},\ \bibinfo {pages} {433} (\bibinfo {year}
  {1903})%
  \bibAnnoteFile{NoStop}{langevin1903}%
\bibitem{sze1981}%
  \BibitemOpen
  \bibfield{author}{%
  \bibinfo {author} {\bibfnamefont{S.~M.}\ \bibnamefont{Sze}},\ }%
  \emph{\bibinfo {title} {Physics of semiconductor devices, {S}econd edition}}\
  (\bibinfo {publisher} {John Wiley \& Sons, Inc.},\ \bibinfo {year} {1981})%
  \bibAnnoteFile{NoStop}{sze1981}%
\bibitem{kemerink2006}%
  \BibitemOpen
  \bibfield{author}{%
  \bibinfo {author} {\bibfnamefont{M.}~\bibnamefont{Kemerink}}, \bibinfo
  {author} {\bibfnamefont{J.~M.}\ \bibnamefont{Kramer}}, \bibinfo {author}
  {\bibfnamefont{H.~H.~P.}\ \bibnamefont{Gommans}},\ and\ \bibinfo {author}
  {\bibfnamefont{R.~A.~J.}\ \bibnamefont{Janssen}},\ }%
  \bibfield{journal}{%
  \Doi{10.1063/1.2205007}{\bibinfo {journal} {Appl. Phys. Lett.}}\ }%
  \textbf{\bibinfo {volume} {88}},\ \bibinfo {pages} {192108} (\bibinfo {year}
  {2006})%
  \bibAnnoteFile{NoStop}{kemerink2006}%
\bibitem{blood}%
  \BibitemOpen
  \bibfield{author}{%
  \bibinfo {author} {\bibfnamefont{P.}~\bibnamefont{Blood}}\ and\ \bibinfo
  {author} {\bibfnamefont{J.~W.}\ \bibnamefont{Orton}},\ }%
  \emph{\bibinfo {title} {{The Electrical Characterization of Semiconductors:
  Majority Carriers and Electron States (Techniques of Physics)}}}\ (\bibinfo
  {publisher} {Academic Press},\ \bibinfo {year} {1992})%
  \bibAnnoteFile{NoStop}{blood}%
\bibitem{yakuphanoglu2008}%
  \BibitemOpen
  \bibfield{author}{%
  \bibinfo {author} {\bibfnamefont{F.}~\bibnamefont{Yakuphanoglu}}\ and\
  \bibinfo {author} {\bibfnamefont{B.~F.}\ \bibnamefont{Senkal}},\ }%
  \bibfield{journal}{%
  \bibinfo {journal} {{Polym. Adv. Technol.}}\ }%
  \textbf{\bibinfo {volume} {19}},\ \bibinfo {pages} {1882} (\bibinfo {year}
  {2008})%
  \bibAnnoteFile{NoStop}{yakuphanoglu2008}%
\bibitem{bisquert2008}%
  \BibitemOpen
  \bibfield{author}{%
  \bibinfo {author} {\bibfnamefont{J.}~\bibnamefont{Bisquert}}, \bibinfo
  {author} {\bibfnamefont{G.}~\bibnamefont{Garcia-Belmonte}}, \bibinfo {author}
  {\bibfnamefont{A.}~\bibnamefont{Munar}}, \bibinfo {author}
  {\bibfnamefont{M.}~\bibnamefont{Sessolo}}, \bibinfo {author}
  {\bibfnamefont{A.}~\bibnamefont{Soriano}},\ and\ \bibinfo {author}
  {\bibfnamefont{H.~J.}\ \bibnamefont{Bolink}},\ }%
  \bibfield{journal}{%
  \bibinfo {journal} {Chem. Phys. Lett.}\ }%
  \textbf{\bibinfo {volume} {465}},\ \bibinfo {pages} {57} (\bibinfo {year}
  {2008})%
  \bibAnnoteFile{NoStop}{bisquert2008}%
\bibitem{deibel2009}%
  \BibitemOpen
  \bibfield{author}{%
  \bibinfo {author} {\bibfnamefont{C.}~\bibnamefont{Deibel}}, \bibinfo {author}
  {\bibfnamefont{T.}~\bibnamefont{Strobel}},\ and\ \bibinfo {author}
  {\bibfnamefont{V.}~\bibnamefont{Dyakonov}},\ }%
  \bibfield{journal}{%
  \bibinfo {journal} {Phys. Rev. Lett.}\ }%
  \textbf{\bibinfo {volume} {103}},\ \bibinfo {pages} {036402} (\bibinfo {year}
  {2009})%
  \bibAnnoteFile{NoStop}{deibel2009}%
\bibitem{pivrikas2005}%
  \BibitemOpen
  \bibfield{author}{%
  \bibinfo {author} {\bibfnamefont{A.}~\bibnamefont{Pivrikas}}, \bibinfo
  {author} {\bibfnamefont{R.}~\bibnamefont{{\"O}sterbacka}}, \bibinfo {author}
  {\bibfnamefont{G.}~\bibnamefont{Ju{\v{s}}ka}}, \bibinfo {author}
  {\bibfnamefont{K.}~\bibnamefont{Arlauskas}},\ and\ \bibinfo {author}
  {\bibfnamefont{H.}~\bibnamefont{Stubb}},\ }%
  \bibfield{journal}{%
  \bibinfo {journal} {Synth. Met.}\ }%
  \textbf{\bibinfo {volume} {155}},\ \bibinfo {pages} {242} (\bibinfo {year}
  {2005})%
  \bibAnnoteFile{NoStop}{pivrikas2005}%
\bibitem{deibel2008b}%
  \BibitemOpen
  \bibfield{author}{%
  \bibinfo {author} {\bibfnamefont{C.}~\bibnamefont{Deibel}}, \bibinfo {author}
  {\bibfnamefont{A.}~\bibnamefont{Baumann}},\ and\ \bibinfo {author}
  {\bibfnamefont{V.}~\bibnamefont{Dyakonov}},\ }%
  \bibfield{journal}{%
  \bibinfo {journal} {Appl. Phys. Lett.}\ }%
  \textbf{\bibinfo {volume} {93}},\ \bibinfo {pages} {163303} (\bibinfo {year}
  {2008})%
  \bibAnnoteFile{NoStop}{deibel2008b}%
\bibitem{deibel2009prb}%
  \BibitemOpen
  \bibfield{author}{%
  \bibinfo {author} {\bibfnamefont{C.}~\bibnamefont{Deibel}}, \bibinfo {author}
  {\bibfnamefont{A.}~\bibnamefont{Wagenpfahl}},\ and\ \bibinfo {author}
  {\bibfnamefont{V.}~\bibnamefont{Dyakonov}},\ }%
  \bibfield{journal}{%
  \bibinfo {journal} {Phys. Rev. B}\ }%
  \textbf{\bibinfo {volume} {80}},\ \bibinfo {pages} {075203} (\bibinfo {year}
  {2009})%
  \bibAnnoteFile{NoStop}{deibel2009prb}%
\end{thebibliography}
\end{document}